# OPPORTUNISTIC SCHEDULING IN UNDERLAY COGNITIVE RADIO BASED SYSTEMS: USER SELECTION PROBABILITY ANALYSIS

*Neeraj Varshney*[‡] *Prabhat K. Sharma*[†] *Mohamed Slim Alouini*[§]

[‡]Department of Electrical Engineering and Computer Science,
Syracuse University, Syracuse, NY 13244, USA (e-mail: nvarshne@syr.edu).
[†]Department of Electronics and Communication Engineering,
VNIT, Nagpur 440010, India (e-mail: prabhatsharma@ece.vnit.ac.in).
[§]Computer, Electrical and Mathematical Science and Engineering Division,
KAUST, Thuwal 23955-6900, Saudi Arabia (e-mail: slim.alouini@kaust.edu.sa).

## ABSTRACT

In this paper, an underlay cognitive radio (CR) system is considered with multiple cognitive or secondary users contending to transmit their information to the cognitive destination (e.g., eNodeB) using the spectral resource of a primary user. The novel closed-form expressions are derived for the selection probabilities of cognitive users with opportunistic scheduling wherein an optimal metric is employed for opportunistic transmission. The analytical results corroborated by the Monte Carlo simulations, can be used to demonstrate the fairness achieved in opportunistic scheduling. It is shown that the fairness in terms of equal chance for transmission amongst all cognitive users can only be seen for the scenarios when the fraction of distances between the cognitive transmitter and cognitive receiver, and cognitive transmitter and primary receiver is identical for each of the cognitive transmitters.

## 1. INTRODUCTION

Cognitive radio (CR) based schemes originally proposed in [1], have received significant attention in cellular and satellite-terrestrial networks due to exponentially increasing number of telecom and satellite services consuming the limited spectral resources. The idea is to enable opportunistic access of the 'licensed' spectrum originally allocated to the incumbent or primary users (PUs) by 'unlicensed' cognitive or secondary users (SUs) in any of the three possible paradigms known as underlay, interweave, and overlay. In contrast to overlay and interweave approaches, underlay is one of the most commonly used schemes where careful power control enables the coexistence of PUs and SUs in same licensed spectrum [2]. However, the power regulation constraint significantly affects the performance of cognitive transmission, especially for the scenarios when the SU transmitter (SU-TX) is located close to the PU receiver (PU-RX). This work, therefore, considers the opportunistic scheduling based underlay cognitive system, where one out of multiple SUs opportunistically access the licensed spectrum of a PU at a given time to enhance the performance of cognitive transmission. These systems have received significant prominence in recent times due to the ever increasing number of users and data hungry applications.

Several works such as [3–6] and the references therein, have recently analyzed the performance of opportunistic scheduling based underlay CR networks in terms of outage probability, symbol error rate (SER), etc. However, to the best of our knowledge, none of these works considered the selection probability analysis of the SU-TXs employing an optimal metric for the opportunistic scheduling. These selection probabilities are required to demonstrate the fairness of opportunistic scheduling schemes with multiple users. Motivated through this fact, this work derives the novel closed-form expressions for the selection probabilities of SU-TXs in an underlay multiuser SU network, and also demonstrates the impact of location of SU-TX and PU-RX on the selection probabilities.

## 2. SYSTEM MODEL

Consider an underlay based CR scenario with opportunistic scheduling, where $K$ SU-TXs request to reuse the licensed spectral resource of a primary user. However, one out of $K$ SU-TXs is opportunistically selected at a time to transmit the information to the cognitive destination (e.g., eNodeB) in the presence of primary communication. It is worth noting that since only one SU-TX transmits at a given time using the spectral resource of the PU-RX, the eNodeB only experiences interference from the primary transmitter (PU-TX). Moreover, to limit the interference at the PU-RX, the selected SU-TX adaptively controls its transmit power using the fixed interference power constraint. The received signal $y_{SD}^{(k)}$ at the eNodeB corresponding to the transmission of a modulated symbol $x_S^{(k)}$ by the $k$th SU-TX is given as

$$y_{SD}^{(k)} = \sqrt{P_S^{(k)}} h_{SD}^{(k)} x_S^{(k)} + \sqrt{P_U} h_{PD} x_P + w_{SD}^{(k)}, \quad (1)$$

where $h_{SD}^{(k)}$ is the channel coefficient for the $k$th SU-TX and eNodeB link, $h_{PD}$ is the channel coefficient for the eNodeB and PU-TX link, $w_{SD}^{(k)}$ denotes additive white Gaussian noise with power $\eta_0$, $x_P$ represents the transmitted symbol by the PU-TX, and $P_S^{(k)}$, $P_U$ are the transmit powers at the $k$th SU-TX and PU-TX, respectively. To prevent interference at the PU-RX, the transmit power $P_S^{(k)}$ at $k$th SU-TX must satisfy

$$P_S^{(k)} = \begin{cases} P_M & \text{if } |h_{SP}^{(k)}|^2 \leq \frac{P_A}{P_M}, \\ \frac{P_A}{|h_{SP}^{(k)}|^2} & \text{if } |h_{SP}^{(k)}|^2 > \frac{P_A}{P_M}, \end{cases} = \min \left\{ P_M, \frac{P_A}{|h_{SP}^{(k)}|^2} \right\},$$

where $h_{SP}^{(k)}$ is the channel coefficient for the $k$th SU-TX and PU-RX link, $P_M$ denotes the maximum transmit power of the SU-TX, and $P_A$ represents the interference threshold at the PU-RX. Similar to existing works [7–9], this work also assumes $P_M >> \frac{P_A}{|h_{SP}^{(k)}|^2}$ for analytical tractability. However, this assumption can be relaxed but at the expense of more involved analytical treatments. Using (1) with $P_S^{(k)} = \frac{P_A}{|h_{SP}^{(k)}|^2}$, the instantaneous SNR at the eNodeB can be obtained as

$$\gamma_{SD}^{(k)} = \frac{P_A |h_{SD}^{(k)}|^2}{(\eta_0 + P_U |h_{PU}|^2) |h_{SP}^{(k)}|^2}. \quad (2)$$

Using the above expression, the optimal metric $\beta^*$ for opportunistic scheduling of secondary users for transmission is given as[1]

$$\beta^* = \max_{k=1,2,\cdots,K} \left\{ \frac{G_{SD}^{(k)}}{G_{SP}^{(k)}} \right\} = \max_{k=1,2,\cdots,K} \left\{ G_S^{(k)} \right\}, \quad (3)$$

where $G_S^{(k)} = \frac{G_{SD}^{(k)}}{G_{SP}^{(k)}}$ and $G_{SD}^{(k)} = |h_{SD}^{(k)}|^2$, $G_{SP}^{(k)} = |h_{SP}^{(k)}|^2$ denote the gain of the cognitive $k$th SU TX-eNodeB and $k$th SU TX-PU-RX links, respectively.

## 3. SELECTION PROBABILITY ANALYSIS

For the selection metric in (3), the probability of the $k$th SU-TX being selected for transmission can be obtained as

$$\begin{aligned} \Pr(k\text{th user}) &= \Pr\left( G_S^{(k)} \geq \max_{\substack{l=1,2,\cdots,K \\ l \neq k}} \left\{ G_S^{(l)} \right\} \right) \\ &= \Pr\left( G_S^{(k)} \geq \widetilde{G}_S \right), \end{aligned} \quad (4)$$

[1]The opportunistic scheduling scheme considered in this work does not require any information about the primary user interference.

where $\widetilde{G}_S$ is defined as, $\widetilde{G}_S \triangleq \max_{\substack{l=1,2,\cdots,K \\ l \neq k}} \left\{ G_S^{(l)} \right\}$. The above expression can be further solved as

$$\begin{aligned} \Pr(k\text{th user}) &= \int_0^\infty \Pr(\widetilde{G}_S \leq y) f_{G_S^{(k)}}(y) dy \\ &= \int_0^\infty F_{\widetilde{G}_S}(y) f_{G_S^{(k)}}(y) dy, \end{aligned} \quad (5)$$

where $F_{\widetilde{G}_S}(y)$ and $f_{G_S^{(k)}}(y)$ denote the CDF of $\widetilde{G}_S$ and PDF of $G_S^{(k)}$, respectively. The CDF $F_{\widetilde{G}_S}(y)$ considering Rayleigh fading links between $k$th SU-TX and eNodeB, and $k$th SU-TX and PU-RX with average gain $\delta_{SD,k}^2$ and $\delta_{SP,k}^2$ respectively, can be derived as

$$F_{\widetilde{G}_S}(y) = \Pr\left( \max_{\substack{l=1,2,\cdots,K \\ l \neq k}} \left\{ G_S^{(l)} \right\} \leq y \right) = \prod_{\substack{l=1 \\ l \neq k}}^{K} F_{G_S^{(l)}}(y), \quad (6)$$

where the CDF $F_{G_S^{(l)}}(y)$ can be solved as

$$F_{G_S^{(l)}}(y) = \Pr\left( \frac{G_{SD}^{(l)}}{G_{SP}^{(l)}} \leq y \right) = \int_0^\infty F_{G_{SD}^{(l)}}(xy) f_{G_{SP}^{(l)}}(x) dx.$$

Substituting $f_{G_{SP}^{(l)}}(x) = \frac{1}{\delta_{SP,l}^2} \exp\left(-\frac{x}{\delta_{SP,l}^2}\right)$ and $F_{G_{SD}^{(l)}}(xy) = 1 - \exp\left(-\frac{xy}{\delta_{SD,l}^2}\right)$, the above integral can be readily solved as

$$F_{G_S^{(l)}}(y) = 1 - \left( 1 + \frac{\delta_{SP,l}^2}{\delta_{SD,l}^2} y \right)^{-1}. \quad (7)$$

Using the above expression in (6), the CDF $F_{\widetilde{G}_S}(y)$ can be written as

$$F_{\widetilde{G}_S}(y) = \prod_{\substack{l=1 \\ l \neq k}}^{K} \left[ 1 - \left( 1 + \frac{\delta_{SP,l}^2}{\delta_{SD,l}^2} y \right)^{-1} \right]. \quad (8)$$

Further, by differentiating $F_{G_S^{(k)}}(y) = 1 - \left( 1 + \frac{\delta_{SP,k}^2}{\delta_{SD,k}^2} y \right)^{-1}$, one can readily derive the PDF $f_{G_S^{(k)}}(y)$ of $G_S^{(k)}$ as

$$f_{G_S^{(k)}}(y) = \frac{\delta_{SP,k}^2}{\delta_{SD,k}^2} \left( 1 + \frac{\delta_{SP,k}^2}{\delta_{SD,k}^2} y \right)^{-2}. \quad (9)$$

It is worth mentioning that the integral expression (5) for the probability of the $k$th SU-TX being selected out of total $K$ users is analytical intractable due to product terms in (8). Therefore, to develop several interesting insights[2] into the selection probabilities, one can solve the integral in (5) as follows.

[2]For an arbitrary value of $K$, it is difficult to get a general expression even after applying the extreme value theorem with $K$ tends to infinity. However, for specific values of $K$, i.e., $K$= 2 or 3, the selection probability of each user can be analytically obtained in closed-form. It is also important to note that higher values of $K$ do not add any new insights.

### 3.1. $K = 2$ Cognitive Users

The probability that the 1st SU-TX is selected for transmission can be obtained as

$$\Pr(\text{1st user}) = \int_0^\infty F_{G_S^{(2)}}(y) f_{G_S^{(1)}}(y) dy. \quad (10)$$

Substituting $F_{G_S^{(2)}}(y) = 1 - \left(1 + \frac{\delta_{SP,2}^2}{\delta_{SD,2}^2} y\right)^{-1}$ and $f_{G_S^{(1)}}(y) = \frac{\delta_{SP,1}^2}{\delta_{SD,1}^2}\left(1 + \frac{\delta_{SP,1}^2}{\delta_{SD,1}^2} y\right)^{-2}$, the above expression can be written as

$$\Pr(\text{1st user}) = \frac{\delta_{SP,1}^2}{\delta_{SD,1}^2}\left[\int_0^\infty \left(1+\frac{\delta_{SP,1}^2}{\delta_{SD,1}^2} y\right)^{-2} dy - \int_0^\infty \left(1+\frac{\delta_{SP,1}^2}{\delta_{SD,1}^2} y\right)^{-2}\left(1+\frac{\delta_{SP,2}^2}{\delta_{SD,2}^2} y\right)^{-1} dy\right]. \quad (11)$$

Further, using the integral identities

$$\int_0^\infty (1+ay)^{-2} dy = \frac{1}{a}, \quad (12)$$

$$\int_0^\infty (1+ay)^{-2}(1+by)^{-1} dy = \frac{1}{a-b}\left[1 - \frac{b\log(a)}{a-b} + \frac{b\log(b)}{a-b}\right],\ a \neq b, \quad (13)$$

the above expression can be solved to yield the final expression for the selection probability of 1st SU-TX as

$$\Pr(\text{1st user}) = 1 - \frac{\alpha_1}{\alpha_1 - \alpha_2}\left[1 - \frac{\alpha_2\log(\alpha_1)}{\alpha_1 - \alpha_2} + \frac{\alpha_2\log(\alpha_2)}{\alpha_1 - \alpha_2}\right], \quad (14)$$

where $\alpha_1$ and $\alpha_2$ are defined as, $\alpha_1 = \frac{\delta_{SP,1}^2}{\delta_{SD,1}^2}$ and $\alpha_2 = \frac{\delta_{SP,2}^2}{\delta_{SD,2}^2}$, respectively. Subsequently, the selection probability of the 2nd SU-TX can be obtained as

$$\Pr(\text{2nd user}) = 1 - \Pr(\text{1st user}). \quad (15)$$

### 3.2. $K = 3$ Cognitive Users

Under the scenario with $K$=3 SU-TXs, the probability of 1st SU-TX is being selected for transmission can be derived as

$$\Pr(\text{1st user}) = \int_0^\infty F_{G_S^{(2)}}(y) F_{G_S^{(3)}}(y) f_{G_S^{(1)}}(y) dy. \quad (16)$$

Substituting $F_{G_S^{(2)}}(y) = 1 - \left(1 + \frac{\delta_{SP,2}^2}{\delta_{SD,2}^2} y\right)^{-1}$, $F_{G_S^{(3)}}(y) = 1 - \left(1 + \frac{\delta_{SP,3}^2}{\delta_{SD,3}^2} y\right)^{-1}$, and $f_{G_S^{(1)}}(y) = \frac{\delta_{SP,1}^2}{\delta_{SD,1}^2}\left(1 + \frac{\delta_{SP,1}^2}{\delta_{SD,1}^2} y\right)^{-2}$, the above expression can be written as

$$\begin{aligned}\Pr(\text{1st user}) = \frac{\delta_{SP,1}^2}{\delta_{SD,1}^2}\Bigg[&\int_0^\infty \left(1+\frac{\delta_{SP,1}^2}{\delta_{SD,1}^2} y\right)^{-2} dy - \int_0^\infty \left(1+\frac{\delta_{SP,1}^2}{\delta_{SD,1}^2} y\right)^{-2}\left(1+\frac{\delta_{SP,2}^2}{\delta_{SD,2}^2} y\right)^{-1} dy \\ &- \int_0^\infty \left(1+\frac{\delta_{SP,1}^2}{\delta_{SD,1}^2} y\right)^{-2}\left(1+\frac{\delta_{SP,3}^2}{\delta_{SD,3}^2} y\right)^{-1} dy \\ &+ \int_0^\infty \left(1+\frac{\delta_{SP,1}^2}{\delta_{SD,1}^2} y\right)^{-2}\left(1+\frac{\delta_{SP,2}^2}{\delta_{SD,2}^2} y\right)^{-1} \times \left(1+\frac{\delta_{SP,3}^2}{\delta_{SD,3}^2} y\right)^{-1} dy\Bigg]. \end{aligned} \quad (17)$$

Further, using the integral identities (12) and (13) along with

$$\begin{aligned}\int_0^\infty (1+ay)^{-2}(1+by)^{-1}(1+cy)^{-1} dy &= \frac{a}{(a-b)(a-c)} - \frac{a(ab+ac-2bc)\log(a)}{(a-b)^2(a-c)^2} \\ &+ \frac{b^2\log(b)}{(a-b)^2(b-c)} + \frac{c^2\log(c)}{(a-c)^2(-b+c)}, \end{aligned} \quad (18)$$

where $a \neq b \neq c$, the above expression can be solved to yield the final expression for Pr(1st user) as

$$\begin{aligned}\Pr(\text{1st user}) = 1 &- \frac{\alpha_1}{\alpha_1 - \alpha_2}\left[1 - \frac{\alpha_2\log(\alpha_1)}{\alpha_1 - \alpha_2} + \frac{\alpha_2\log(\alpha_2)}{\alpha_1 - \alpha_2}\right] \\ &- \frac{\alpha_1}{\alpha_1 - \alpha_3}\left[1 - \frac{\alpha_3\log(\alpha_1)}{\alpha_1 - \alpha_3} + \frac{\alpha_3\log(\alpha_3)}{\alpha_1 - \alpha_3}\right] \\ &+ \frac{\alpha_1^2}{(\alpha_1-\alpha_2)(\alpha_1-\alpha_3)} - \frac{\alpha_1^2(\alpha_1\alpha_2+\alpha_1\alpha_3-2\alpha_2\alpha_3)\log(\alpha_1)}{(\alpha_1-\alpha_2)^2(\alpha_1-\alpha_3)^2} \\ &+ \frac{\alpha_1\alpha_2^2\log(\alpha_2)}{(\alpha_1-\alpha_2)^2(\alpha_2-\alpha_3)} + \frac{\alpha_1\alpha_3^2\log(\alpha_3)}{(\alpha_1-\alpha_3)^2(-\alpha_2+\alpha_3)}, \end{aligned} \quad (19)$$

where $\alpha_3 = \frac{\delta_{SP,3}^2}{\delta_{SD,3}^2}$. Similarly, the selection probability of 2nd SU-TX can be obtained as

$$\Pr(\text{2nd user}) = \int_0^\infty F_{G_S^{(1)}}(y) F_{G_S^{(3)}}(y) f_{G_S^{(2)}}(y) dy. \quad (20)$$

Substituting $F_{G_S^{(1)}}(y) = 1 - \left(1 + \frac{\delta_{SP,1}^2}{\delta_{SD,1}^2} y\right)^{-1}$, $F_{G_S^{(3)}}(y) = 1 - \left(1 + \frac{\delta_{SP,3}^2}{\delta_{SD,3}^2} y\right)^{-1}$, and $f_{G_S^{(2)}}(y) = \frac{\delta_{SP,2}^2}{\delta_{SD,2}^2}\left(1 + \frac{\delta_{SP,2}^2}{\delta_{SD,2}^2} y\right)^{-2}$, the above expression can be solved as

$$\begin{aligned}\Pr(\text{2nd user}) = 1 &- \frac{\alpha_2}{\alpha_2 - \alpha_1}\left[1 - \frac{\alpha_1\log(\alpha_2)}{\alpha_2 - \alpha_1} + \frac{\alpha_1\log(\alpha_1)}{\alpha_2 - \alpha_1}\right] \\ &- \frac{\alpha_2}{\alpha_2 - \alpha_3}\left[1 - \frac{\alpha_3\log(\alpha_2)}{\alpha_2 - \alpha_3} + \frac{\alpha_3\log(\alpha_3)}{\alpha_2 - \alpha_3}\right] \\ &+ \frac{\alpha_2\alpha_2}{(\alpha_2-\alpha_1)(\alpha_2-\alpha_3)} - \frac{\alpha_1\alpha_2(\alpha_2\alpha_1+\alpha_2\alpha_3-2\alpha_1\alpha_3)\log(\alpha_2)}{(\alpha_2-\alpha_1)^2(\alpha_2-\alpha_3)^2} \\ &+ \frac{\alpha_1\alpha_1^2\log(\alpha_1)}{(\alpha_2-\alpha_1)^2(\alpha_1-\alpha_3)} + \frac{\alpha_1\alpha_3^2\log(\alpha_3)}{(\alpha_2-\alpha_3)^2(-\alpha_1+\alpha_3)}. \end{aligned} \quad (21)$$

Subsequently, the selection probability of the 3rd SU-TX can be obtained as

$$\Pr(\text{3rd user}) = 1 - \Pr(\text{1st user}) - \Pr(\text{2nd user}). \quad (22)$$

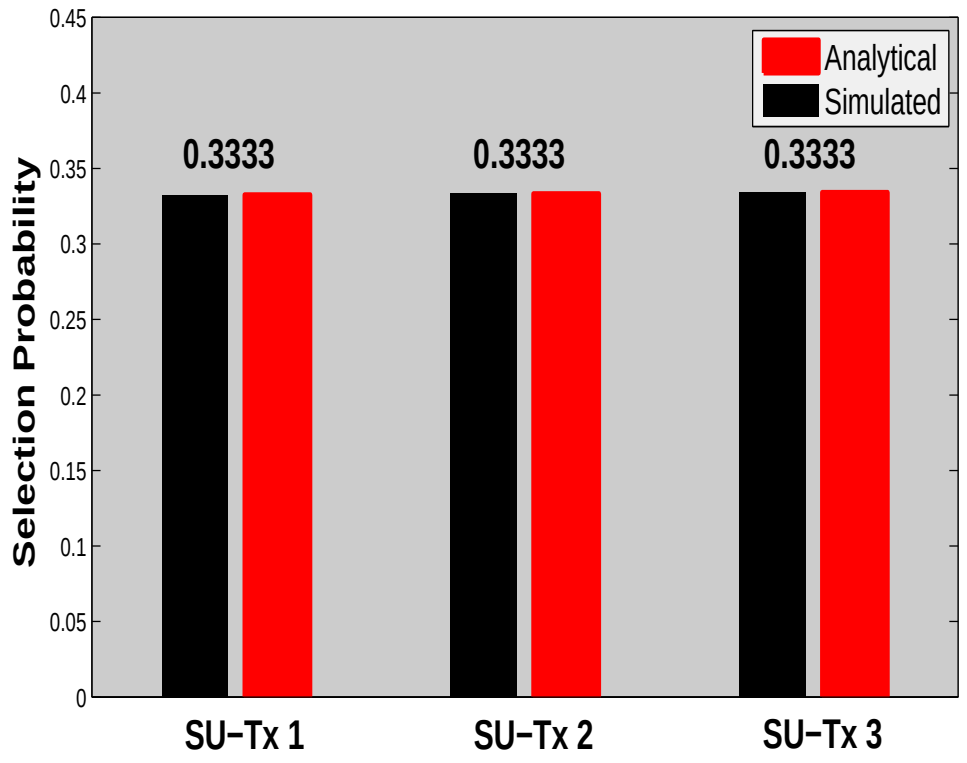

**Fig. 1**. Selection Probabilities of SU-TXs when each SU-TX is located at approximate equal distance from eNodeB and PU-RX, i.e., $d_{SD,k} \approx d_{SP,k} \approx 2 \; \forall k$.

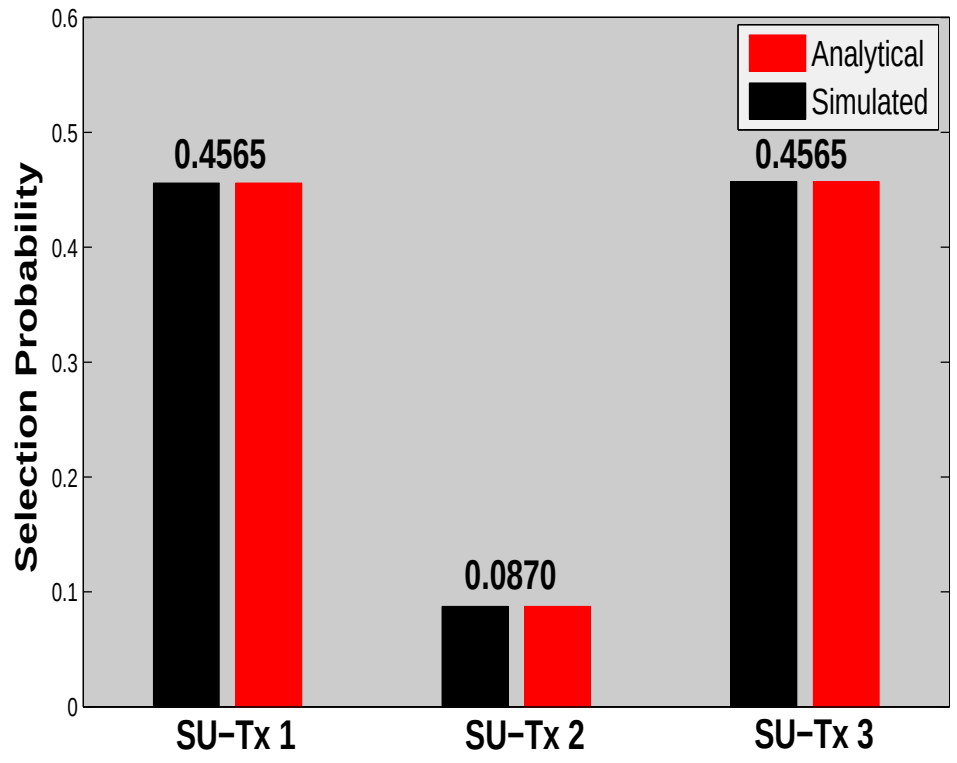

**Fig. 3**. Selection Probabilities of SU-TXs when 2nd SU-TX is close to the PU-RX in comparison to the eNodeB, i.e., $d_{SP,k} \approx d_{SD,k} \approx 2$, $\forall k$ except $d_{SP,2} \approx 1$.

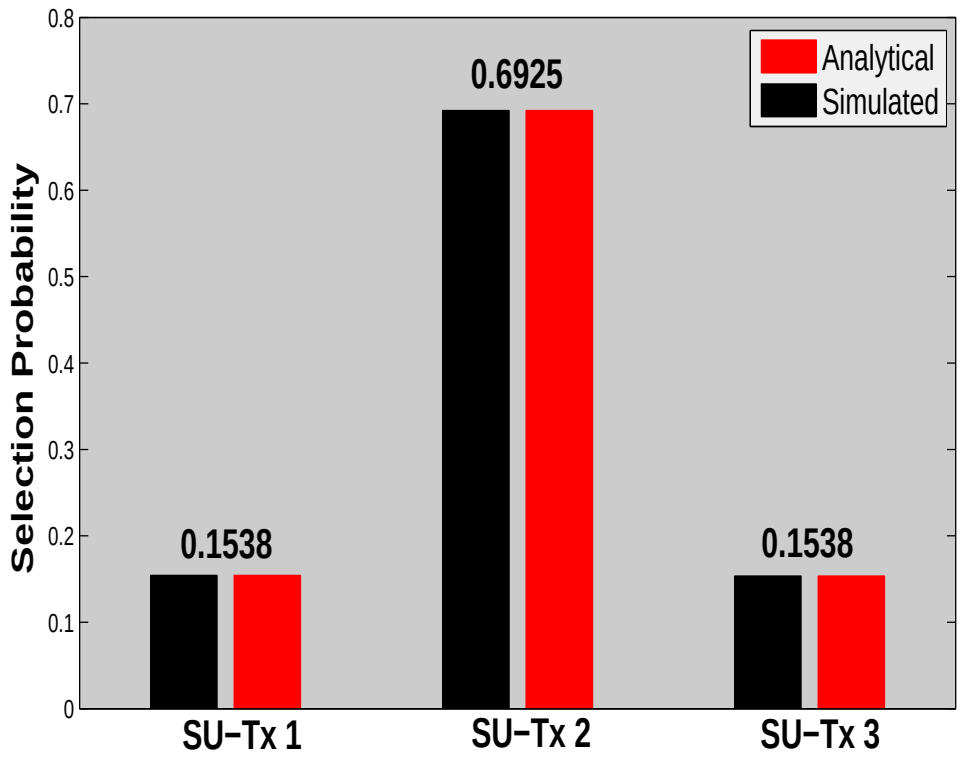

**Fig. 2**. Selection Probabilities of SU-TXs when 2nd SU-TX is closer to the eNodeB than the PU-RX, i.e., $d_{SP,k} \approx d_{SD,k} \approx 2$, $\forall k$ except $d_{SD,2} \approx 1$.

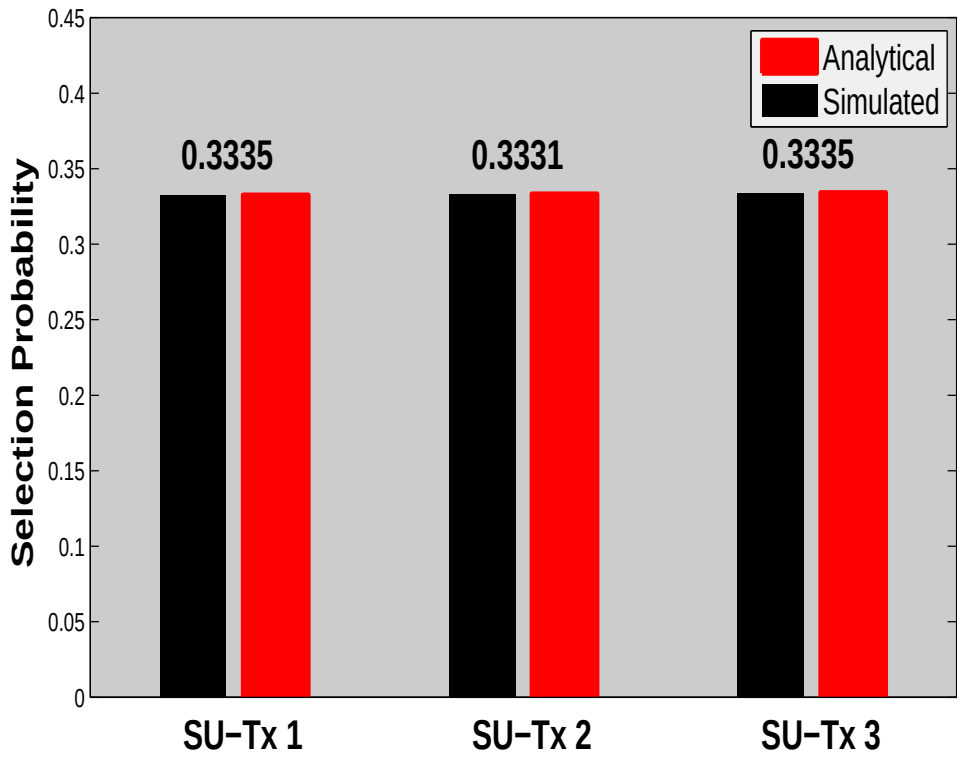

**Fig. 4**. Selection Probabilities of SU-TXs when 2nd SU-TX is located close to the $R$ in comparison with the other SU-TXs and also have equal distance from eNodeB and PU-RX, i.e., $d_{SP,k} \approx d_{SD,k} \approx 2$, $\forall k$ except $d_{SD,2} \approx d_{SP,2} \approx 1$.

## 4. SIMULATION RESULTS

This section presents simulation results to develop several interesting insights into the selection probabilities of SU-TXs. For simulation purposes, we consider the presence of $K = 3$ SU-TXs, and obtain the average channel gains $\delta^2_{SD,k}$ and $\delta^2_{SP,k}$ as, $\delta^2_{SD,k} = d^{-\beta}_{SD,k}$ and $\delta^2_{SP,k} = d^{-\beta}_{SP,k}$, respectively. Here $\beta = 3$ is the path loss exponent, and $d_{SD,k}$, $d_{SP,k}$ denote the distances between the $k$th SU-TX and eNodeB, and the $k$th SU-TX and PU-RX, respectively. Fig. 1 shows the selection probabilities of SU-TXs when each SU-TX is located at approximately[3] equal distances from eNodeB and PU-RX, i.e., $d_{SD,k} \approx d_{SP,k} \approx 2 \; \forall k$. It can be observed that for the scenarios when each SU-TX is located at equal distance from eNodeB and PU-RX, each SU-TX has equal probability of being selected for transmission.

For the scenario, when 2nd SU-TX is closer to the eNodeB than the PU-RX, i.e., $d_{SP,k} \approx d_{SD,k} \approx 2$, $\forall k$ except $d_{SD,2} \approx 1$ as shown in Fig. 2, the system enhances the performance by choosing 2nd SU-TX approximately 69% of the times. Consequently, the selection probability of other SU-TXs significantly reduces to 0.1538.

On the other hand, when 2nd SU-TX is close to the PU-RX in comparison to the eNodeB, i.e., $d_{SP,k} \approx d_{SD,k} \approx 2$, $\forall k$ except $d_{SP,2} \approx 1$, the probability of choosing 2nd SU-TX for transmission reduces to approximately 9% that in turns increases the selection probabilities of other SU-TXs to 0.4565. However, if the 2nd SU-TX is located close to the eNodeB in comparison with the other SU-TXs and also have equal distance from eNodeB and PU-RX, i.e., $d_{SP,k} \approx d_{SD,k} \approx 2$, $\forall k$ except $d_{SD,2} \approx d_{SP,2} \approx 1$, each of the SU-TXs will have approximately equal chance for transmission, as shown in Fig. 4. Therefore, based on above observations, the fairness in terms of equal chance for transmission can only be seen for the scenarios when the fraction of distances i.e., $\frac{d_{SD,k}}{d_{SP,k}}$ is identical for each SU-TX.

[3] The integral in (13) and (18) are solved for the scenario where $a \neq b \neq c$. Therefore, for simulation purposes, we considered $d_{SD,k} \approx d_{SP,k} \approx 2 \; \forall k$, where $d_{SD,1} = 2.002$, $d_{SD,2} = 2.004$, $d_{SD,3} = 2.006$, $d_{SP,1} = 2.001$, $d_{SP,2} = 2.003$, and $d_{SP,3} = 2.005$.